\documentclass[aip,jap,reprint,floatfix,showpacs,amsmath,amssymb]{revtex4-1}
\usepackage{graphicx}
\usepackage{amssymb}
\usepackage{epsfig}
\usepackage{epsf}
\usepackage[usenames]{color}

\usepackage{graphicx}
\usepackage{subfigure}
\def\ba{\begin{eqnarray}}
\def\ea{\end{eqnarray}}
\def\be{\begin{equation}}
\def\ee{\end{equation}}

\begin{document}

\title{Vortex-antivortex pairs induced by curvature in toroidal nanomagnets}
\author{Smiljan Vojkovic}
\affiliation{Instituto de F\'isica, Pontificia Universidad Cat\'olica de Chile, Campus San Joaqu\'in \\ Av. Vicu\~na Mackena 4860, Santiago, Chile.}
\author{Vagson L. Carvalho-Santos}
\affiliation{Instituto Federal de Educa\c c\~ao, Ci\^encia e Tecnologia Baiano - Campus 
Senhor do Bonfim, \\Km 04 Estrada da Igara, 48970-000 Senhor do Bonfim, Bahia, Brazil}
\author{Jakson M. Fonseca}
\affiliation{Universidade Federal de Vi\c cosa, Departamento de F\'isica, 
Avenida Peter Henry Rolfs s/n, 36570-000, Vi\c cosa, MG, Brasil}
\author{Alvaro S. Nunez}
\affiliation{Departamento de F\'isica, Facultad de Ciencias F\'isicas y 
Matem\'aticas, Universidad de Chile, Casilla 487-3, Santiago, Chile}

%
\begin{abstract}
We show that the curvature of nanomagnets can be used to induce chiral textures in the magnetization field. Among the phenomena related to the interplay between geometry and magnetic behavior at nanomagnets, an effective curvature-induced chiral interaction has been recently predicted. In this work, it is shown that a magnetization configuration consisting of two structures with opposite winding numbers (vortex and antivortex) appear as remanent states in hollow toroidal nanomagnets. It is shown that these topological configurations are a result of a chiral interaction induced by curvature. In this way, the obtained results present a new form to produce stable vortices and antivortices by using nanomagnets with variable curvature.
\end{abstract}
\maketitle
\section{Introduction}
Proper control over magnetization textures is at the heart of a large variety of technological applications. This makes of any new tool to control the fate of magnetization patterns a very interesting avenue for research. In this context, topological spin configurations such as chiral skyrmions, magnetic bubble domains, vortices and antivortices have been widely studied due to the  possibility to produce devices based on magnonics and spintronics technology  \cite{bubble0,bubble1, Skyrmion0,Skyrmion-1,Skyrmion-2,Choe-Science-2004,Hertel-PRL-2007,Xiao-APL-2006,Gliga-PRB-2008,Wang-PRB-2007,Mironov-PRB-2010, Roldan}. For example, skyrmion based ``race-track'' memory devices \cite{Skyrmion-1,Race-Track-Skyrmion} and vortex based transistor  \cite{Kumar-SciRep-2014} have been recently proposed. Chiral skyrmions are two-dimensional topological solitons localized in nanoscale cylindrical regions \cite{Bogdanov1,Bogdanov2,Bogdanov3}. The understanding of the physics of magnetic skyrmions and their applications in potential spintronic devices requires the knowledge of the properties of an isolated skyrmion. In this context, an analytical solution for such isolated skyrmions has been recently obtained and compared with experimental results \cite{Bogdanov4}. Despite vortices being also characterized by a winding number, in 3D systems they do not have topological stability and then, their stabilization in magnetic systems is ensured by subtle competition between the dipolar and exchange energies \cite{Scholz-JMMM-2003}. In this case, vortices can appear as the groundstate of  submicron magnetic elements of different shapes \cite{Shinjo-Science,Shape2,Shape3}. In addition, vortex domain walls are stable configurations in cylindrical nanotubes and their dynamics is under intense investigation \cite{Otalora-APL-2012,Otalora-JMMM-2013}. Unlike skyrmions and vortices, the preparation of a nanostructure that contains a stable antivortex is a challenging task. Indeed, antivortices can appear as an unstable magnetization configuration during a vortex core reversal \cite{Waeyenberge-Nature-2006}. Nevertheless, experimental and theoretical works have reported that a stable single antivortex can appear in asterisk \cite{Shigueto-APL-2002,Gliga-PRB-2008} and cross-like \cite{Mironov-PRB-2010} shaped permalloy particles. In this letter, we provide an additional mechanism for the nucleation, control and transport of chiral textures of the magnetization by appealing to the geometrical curvature of the system.

Curvature has become a cornerstone in the modern description of the behavior of magnetic systems. In fact, a lot of effort has been dedicated to the understanding of the effects that the geometrical properties of a system have on the static and dynamic behavior of the magnetization \cite{Review-Curvature}. For instance, the presence of curvature and torsion in curved wires yield the appearing of a domain wall pinning and a Walker limit \cite{Walker-JAP-1974} during the domain wall motion \cite{Yershov-PRB-2015,Sheka-PRB-2015,Yershov-PRB-2016}. The basics spin wave and domain wall motion physics behind the geometrical confinement of magnetic textures has been investigated in cylindrical nanomagnets \cite{Ferguson-JMMM-2015,Landeros-JMMM-2010,Landeros-JAP-2010,Otalora-JPC-2012} and a curvature-induced asymmetric spin wave has been predicted to appear in the spin wave dispersion along ferromagnetic nanotubes \cite{Otalora-PRL-2016}. As a last example, the coupling between an external magnetic field strength and the surface curvature can lead to the appeareance of a 2$\pi$-skyrmion excitation on magnetic surfaces described by the Heisenberg Hamiltonian \cite{Vagson-JMMM-2015,Vagson-PLA-2012}.

In a recent work, approximating the full extent of the dipolar interaction by a suitably chosen shape anisotropy, Gaididei \textit{et al} \cite{Gaididei-PRL-2014} have obtained a functional to calculate the exchange energy for magnetic shells with an arbitrary geometry in function of its Gaussian and mean curvatures. They have shown that an effective anisotropy and a Dzyaloshinskii-Moriya-like interaction (DM-like interaction) \cite{Dzyaloshinkii,Moryia} are induced by curvature. Such curvature-induced DM-like interaction was firstly predicted in the context of a bending in a magnetic film \cite{Hertel-SPIN-2013}. That is, the bending of the film induces a handedness in the magnetization, which becomes more pronounced as the curvature of the film increases in such way that geometry can break the inversion symmetry and give rise to chiral effects. Pylypovskyi \textit{et al} showed that this effective DM-like interaction is responsible by a chiral symmetry breaking of a domain wall motion in magnetic helices \cite{Pylypovskyi-SRep-2016} and M\"obius ring \cite{Pylypovskyi-PRL-2015}. Based on the above and on the fact that the torus presents a curvature varying from negative (internal border) to positive (external border), with a relatively easy geometrical description, we propose that this shape is ideal to analyze the predicted chiral effects induced by the curvature. In this work it is shown that the variable curvature of the torus yields chiral magnetization patterns with opposite winding numbers (vortex and antivortex) and so, a new way to stabilize vortices and antivortices in magnetic nanoparticles by using curvature effects is proposed. The magnetic properties of toroidal nanomagnets have been previously studied \cite{Vagson-JAP-2010,Smiljan-JAP-2016}. However, the  geometrical parameters considered so far do not allow the appearance of more complex structures during reversal process that could evidence a curvature-induced DM-like interaction. Here, we study hollow toroidal nanoparticles that are expected to display the qualitative features rendered by the curvature-induced DM-like interaction \cite{Gaididei-PRL-2014}.  

\section{Theoretical model}

In general, in a curved surface parametrized by the coordinates $(q_1,q_2)$, we can write the magnetization in an orthogonal curvilinear basis, ($\hat{q_1},\hat{q_2},\hat{n}=\hat{q_1}\times\hat{q_2}$), in the form $\vec{M}=M_S\mathbf{m}$, with $M_S$ being the saturation magnetization and 
\ba\label{DefAngMag}
\mathbf{m}=\hat{n}\cos\Theta+\hat{q_1}\sin\Theta\cos\Phi+\hat{q_2}\sin\Theta
\sin\Phi\,,
\ea
where, $\Theta\equiv\Theta(q_1,q_2)$ and $\Phi\equiv\Phi(q_1,q_2)$ describe the angles of the magnetization vector field in a curvilinear background. In this way, the exchange energy density for an arbitrary curved magnetic shell is explicitly written below \cite{Gaididei-PRL-2014,Sheka-JPA-2015}
\ba\label{ExcInt}
\frac{\mathcal{E}_{ex}}{A}=\left[\sin\Theta(\nabla\Phi-\mathbf{\Omega})-
\cos\Theta\frac{\partial\mathbf\Gamma(\Phi)}{\partial\Phi}\right]^2\nonumber\\
+\left[\nabla\Theta-\mathbf\Gamma(\Phi)\right]^2\,,
\ea
where $A$ is the stifness constant, $\mathbf\Gamma(\Phi)$ is a matrix depending on the Gauss and mean curvatures of the nanomagnet and $\mathbf{\Omega}$ is a modified spin connection, defined as \cite{Gaididei-PRL-2014}
\ba\label{SpinConnectionEq}
\mathbf{\Omega}=\left(\hat{q_1}\cdot\frac{\partial\hat{q_2}}{\partial q_1}\right)\hat{q_1}+\left(\hat{q_1}\cdot\frac{\partial\hat{q_2}}{\partial q_2}\right)\hat{q_2}\,.
\ea 

\begin{figure}
{\includegraphics[scale=0.3]{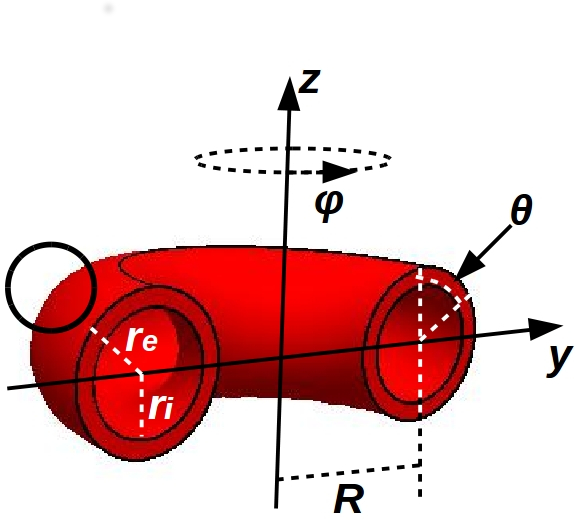}}{\includegraphics[scale=0.2]{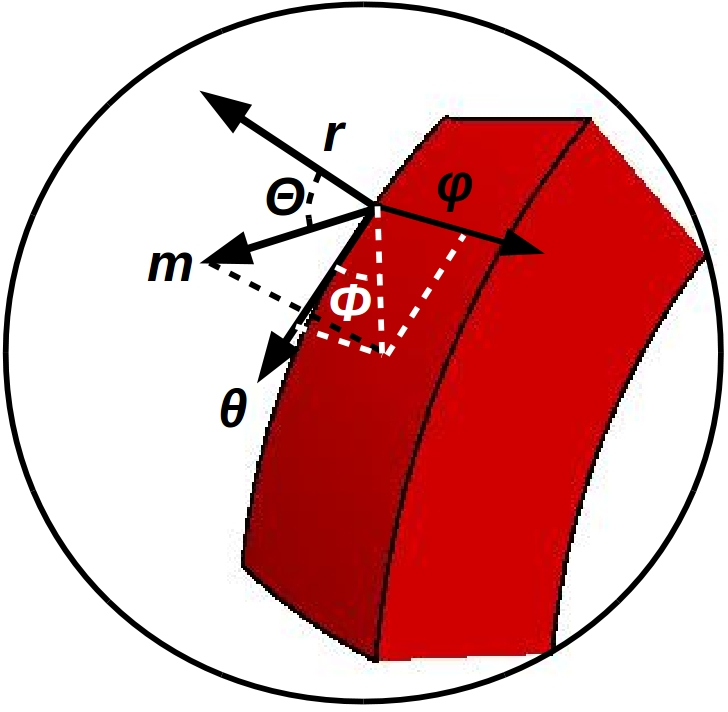}}\caption{Left figue presents a torus section showing the adopted coordinate system describing the torus external surface. Right figure shows a zoom of the region highlighted by a circle with the magnetization parametrization in function of the toroidal unitary vectors ($\mathbf{r},\mathbf{\theta},\mathbf{\varphi}$).}\label{Torus-parameters}
\end{figure}

A torus with genus 1 embedded in a 3D-space can be parametrized by
\ba \label{ParametricEq}
\vec{r}=(R+r\sin\theta)(\hat{x}\,\cos\varphi+\hat{y}\,\sin\varphi)+\hat{z}\,r\cos\theta\,,
\ea 
where $R$ and $r$ are respectively the toroidal and poloidal radii and $\theta$ plays the role of polar angle describing the torus surface (See Fig. \ref{Torus-parameters}). In this case, its Gaussian curvature is evaluated as $K(\theta)=\sin\theta[r(R+r\sin\theta)]^{-1}$ and then, the the curvature of the torus varies from a positive to a negative value along the polar-like angle, that is, $K(\pi/2)=[r(R+r)]^{-1}$ and $K(3\pi/2)=-[r(R-r)]^{-1}$.
From Eq. (\ref{SpinConnectionEq}), the modified spin connection for the toroidal geometry can be calculated from the geometrical parameters defined in Eq. (\ref{ParametricEq}), leading to 
\begin{equation}\label{SPCon}
\mathbf{\Omega_t}(\theta)=-\hat{\varphi}\,\frac{\cos\theta}{R+r\sin\theta}\,.
\end{equation}
It can be noted that the spin connection varies in function of $\theta$, pointing along $-\hat{\varphi}$ for $\theta\in(-\pi/2,\pi/2)$ and $+\hat{\varphi}$ for $\theta\in(\pi/2,3\pi/2)$. In this context, we have that $\mathbf{\Omega_t}(0)=-\hat{\varphi}/R$ and $\mathbf{\Omega_t}(\pi)=\hat{\varphi}/R$. According Gaididei \textit{et al} \cite{Gaididei-PRL-2014}, if magnetostatic energy is approximated by an easy-tangential anisotropy (this is a good approximation for thin shells) and a tangential magnetization configuration ($\Theta=\pi/2$) is considered, the magnetic energy can be split in three components: i) the ``standard'' exchange interaction $\mathcal{E}_{E}=A(\nabla\Phi)^2$ which homogenizes the spatial distribution of the magnetization vector, minimized for $\Phi=constant$; ii) an effective anisotropy, given by $\mathcal{E}_{A}=A\mathbf{\Gamma}^2$; and iii) an effective DM-like interaction, given by $\mathcal{E}_{D}=-2A(\nabla\Phi\cdot\mathbf{\Omega})$. The latter contribution is minimized when the magnetization display inhomogeneous distribution textures. The magnetization ground state can be interpreted as a result of the interplay among dipolar, a curvature-induced magnetic field pointing along the normal direction and these three curvature-induced interactions \cite{Gaididei-PRL-2014}. 

If we consider a general magnetization distribution on a toroidal shell, the curvature-induced DM-like energy density is given by \cite{Sheka-JPA-2015}
\ba\label{DMI-torus}
\frac{\mathcal{E}_{D}}{A}&=&\frac{2}{(R+r\sin\theta)^2}\left[\cos\theta\sin^2\Theta \partial_\varphi\Phi\right.\nonumber\\
&-&\left.\sin\theta(\cos^2\Theta\sin\Phi\partial_\varphi\Theta-
\sin\Theta\cos\Theta\cos\Phi\partial_\varphi\Phi)\right]\nonumber\\
&+&\frac{1}{r^2}(\cos\Phi\partial_\theta\Theta-\cos\Theta\sin\Theta\sin\Phi\partial_\theta\Phi)\,.\,\,\,\,\,
\ea

From Eq. (\ref{DMI-torus}), one can estimate the effective DM-like interaction strength induced by curvature. Indeed, by using a dimensional analysis, we have that $\mathcal{D}/a^2\sim A/R=J/Ra$ and thus, $\mathcal{D}\sim J(a/R)$, where $a$ is the lattice constant and $J$ is the exchange constant. From Eqs. (\ref{DefAngMag}) and (\ref{DMI-torus}), it can be noted that the effective DM-like interaction term to the energy of a in-surface vortex state, given by $\Theta=\pi/2$ and $\Phi=\pi/2$, and a single domain state, represented by $\Theta=\theta+\pi/2$ and $\Phi=\pi/2$ (pointing along $x$-axis direction), is 0 and then, no chiral effects are present for these magnetization fields. Nevertheless, due to its dependence on $\theta$, chiral effects coming from curvature-induced DM-like interaction must be evident when other magnetization configurations appearing in a toroidal nanomagnet are analyzed. Therefore, to study metastable states in a toroidal nanomagnet and analyze the possibility of the appearance of chiral interactions induced by variable curvature, we have performed micromagnetic simulations for hollow Permalloy nanotori and solved the Landau-Lifshitz-Gilbert equation \cite{LLG,Gilbert}
\be\label{LLGEq}
\frac{\partial \mathbf{m}}{\partial t}=\frac{\gamma_0}{\mu_0M}\mathbf{m}\times\frac{\partial\mathcal{E}}{\partial\mathbf{m}}+\alpha\mathbf{m}\times\frac{\partial\mathbf{m}}{\partial t}\,,
\ee
where $\gamma_0=\mu_0|\gamma|=\mu_0g|\mu_B|/\hbar$, with $\gamma$ the gyromagnetic ratio, $M_{_S}$ is the saturation magnetization, $\mathcal{E}$ is the free energy density and $\alpha$ is the dimensionless Gilbert damping parameter. The first term of Eq. (\ref{LLGEq}) describes the precession of the magnetization under the influence of an effective magnetic field $\mathbf{H}_{\text{eff}}=(-1/\mu_0M_{_S})\delta\mathcal{E}/\delta\mathbf{m}$, the second term accounts for the relaxation mechanisms that dissipate energy by making a torque towards the effective field. The effective magnetic field that each magnetic moment experiments is created by the exchange, dipolar and anisotropy interactions, as well as the external magnetic field.

\section{Results}

\begin{figure}
{\includegraphics[scale=0.2]{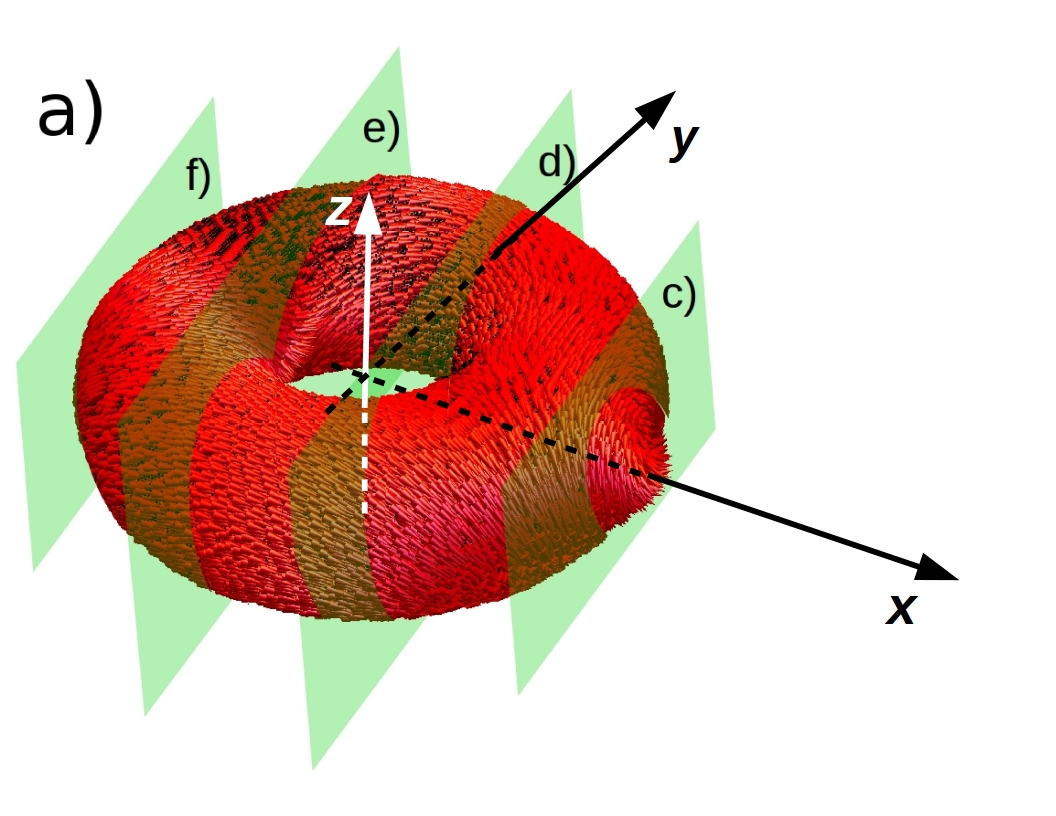}}\\{
\includegraphics[scale=0.14]{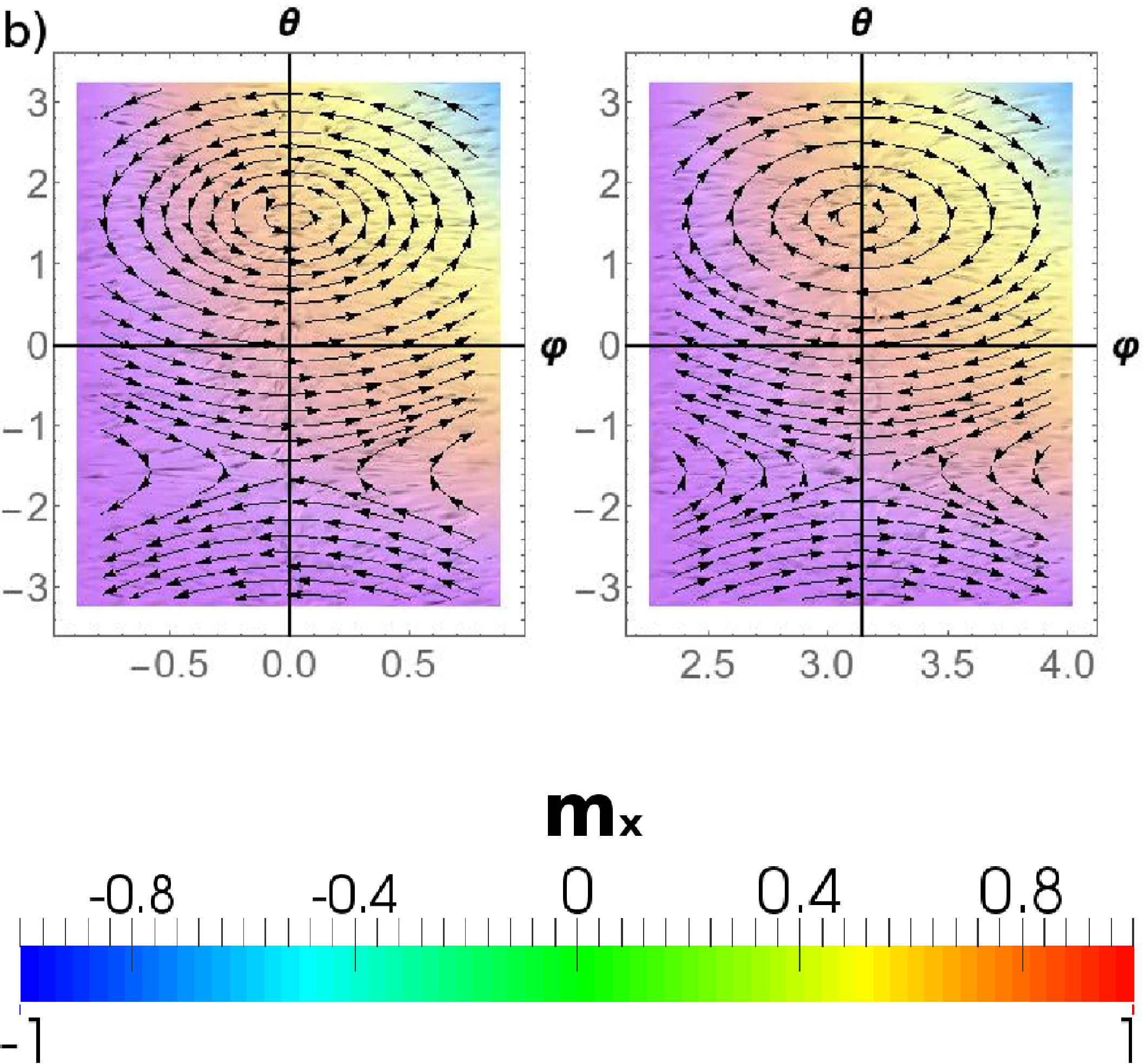}}\\
{\includegraphics[scale=0.072]{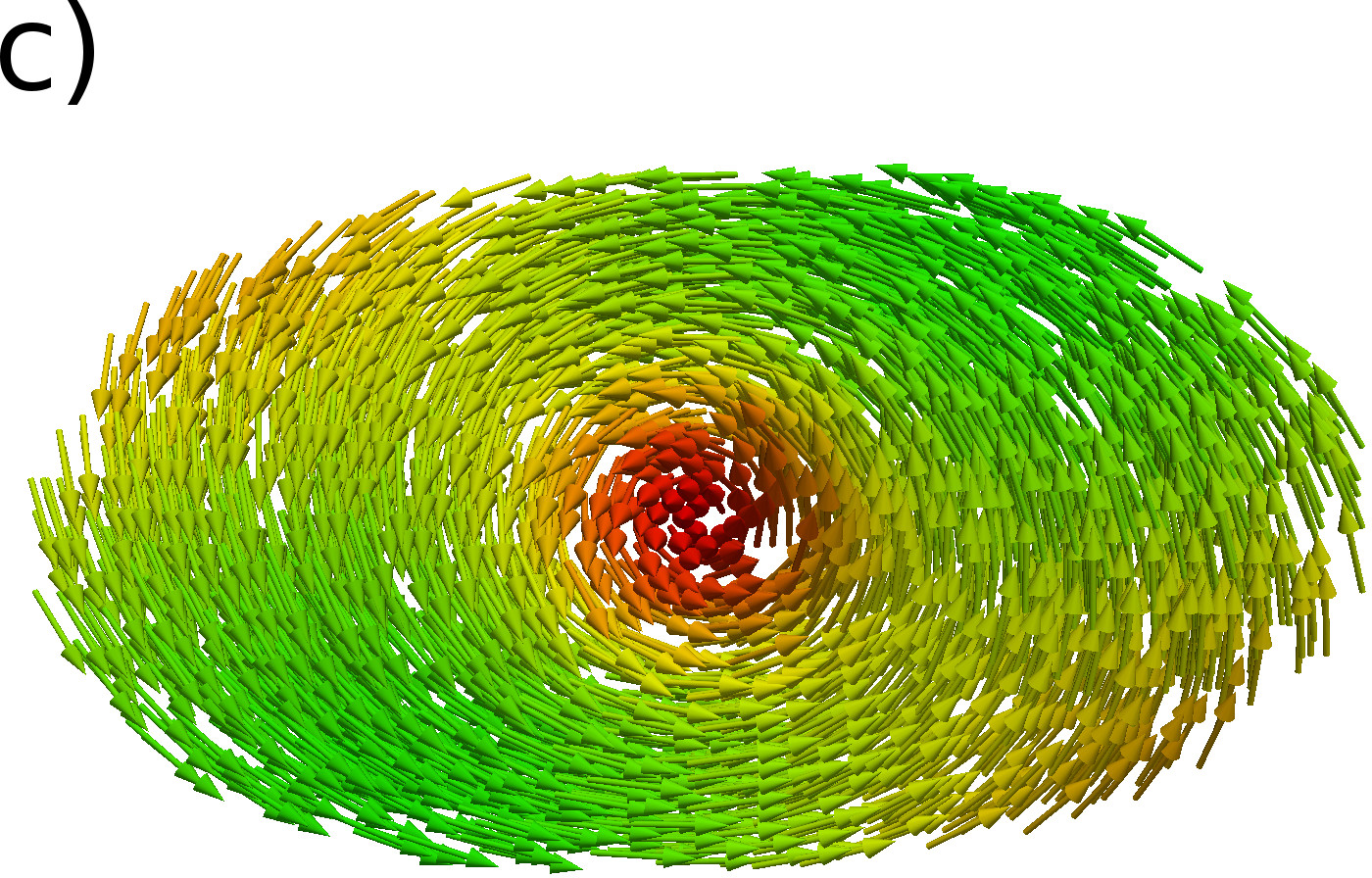}}\hspace{1em}{\includegraphics[scale=0.072]{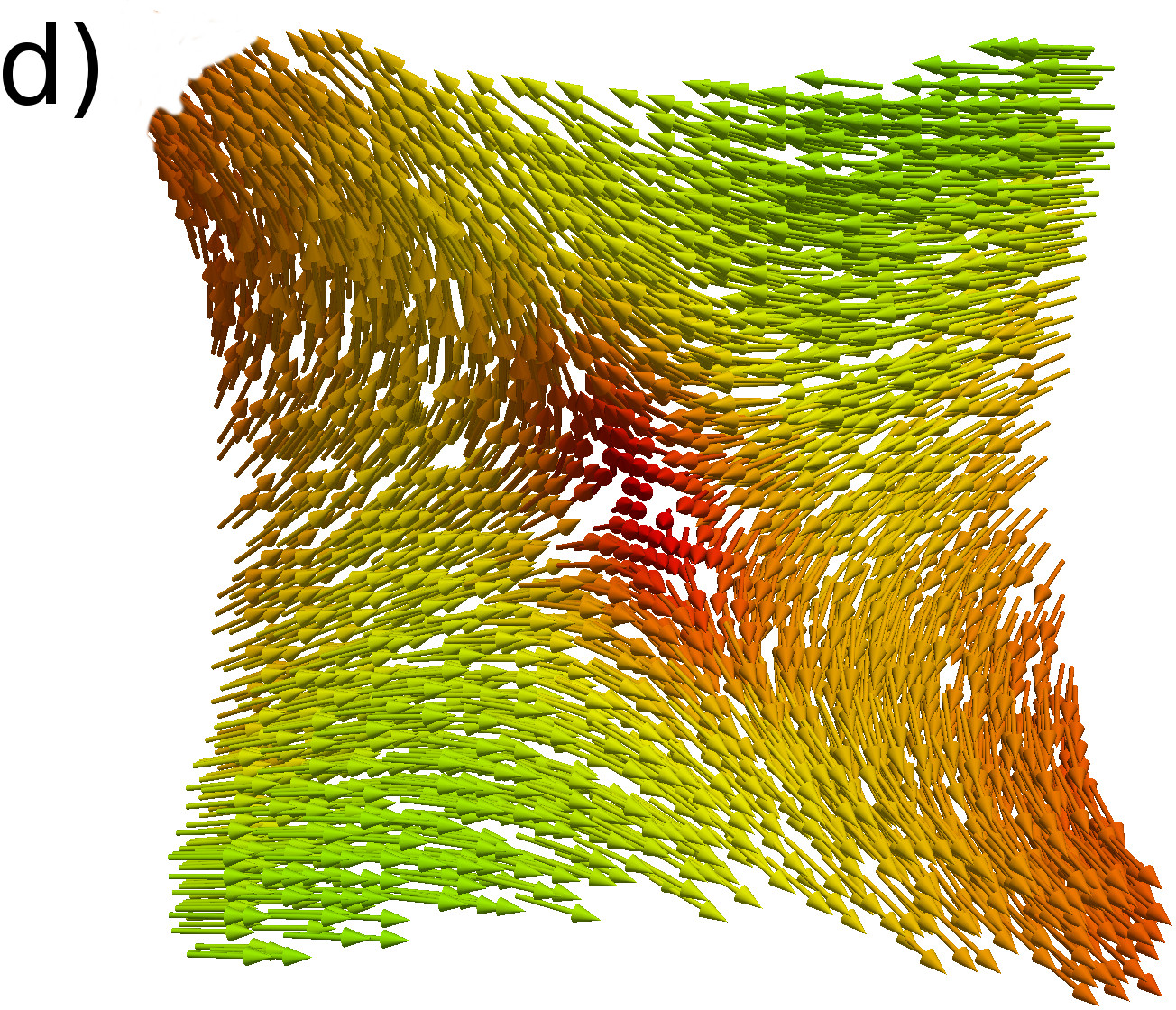}}\\{\includegraphics[scale=0.08]{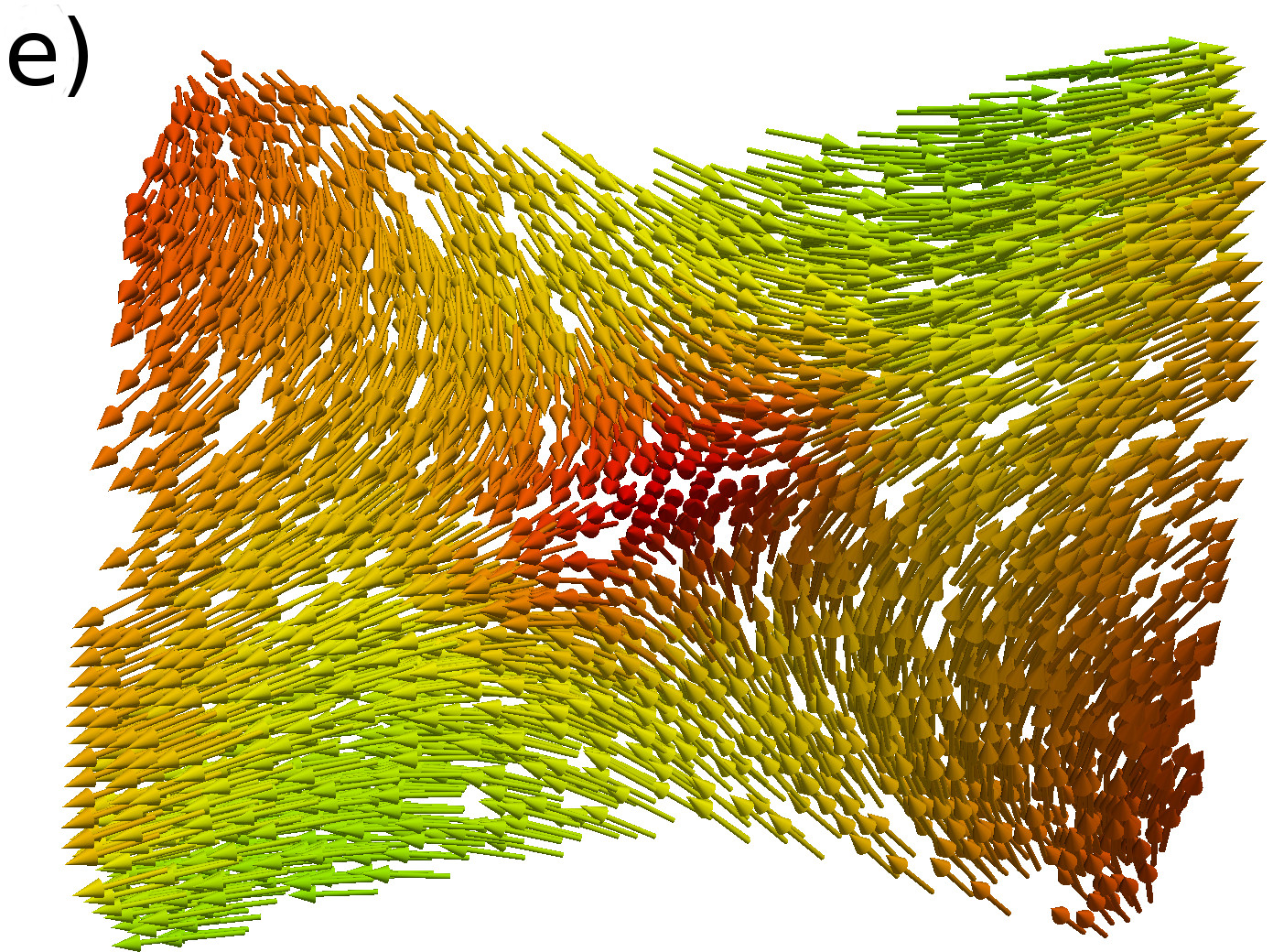}}\hspace{1em}{\includegraphics[scale=0.08]{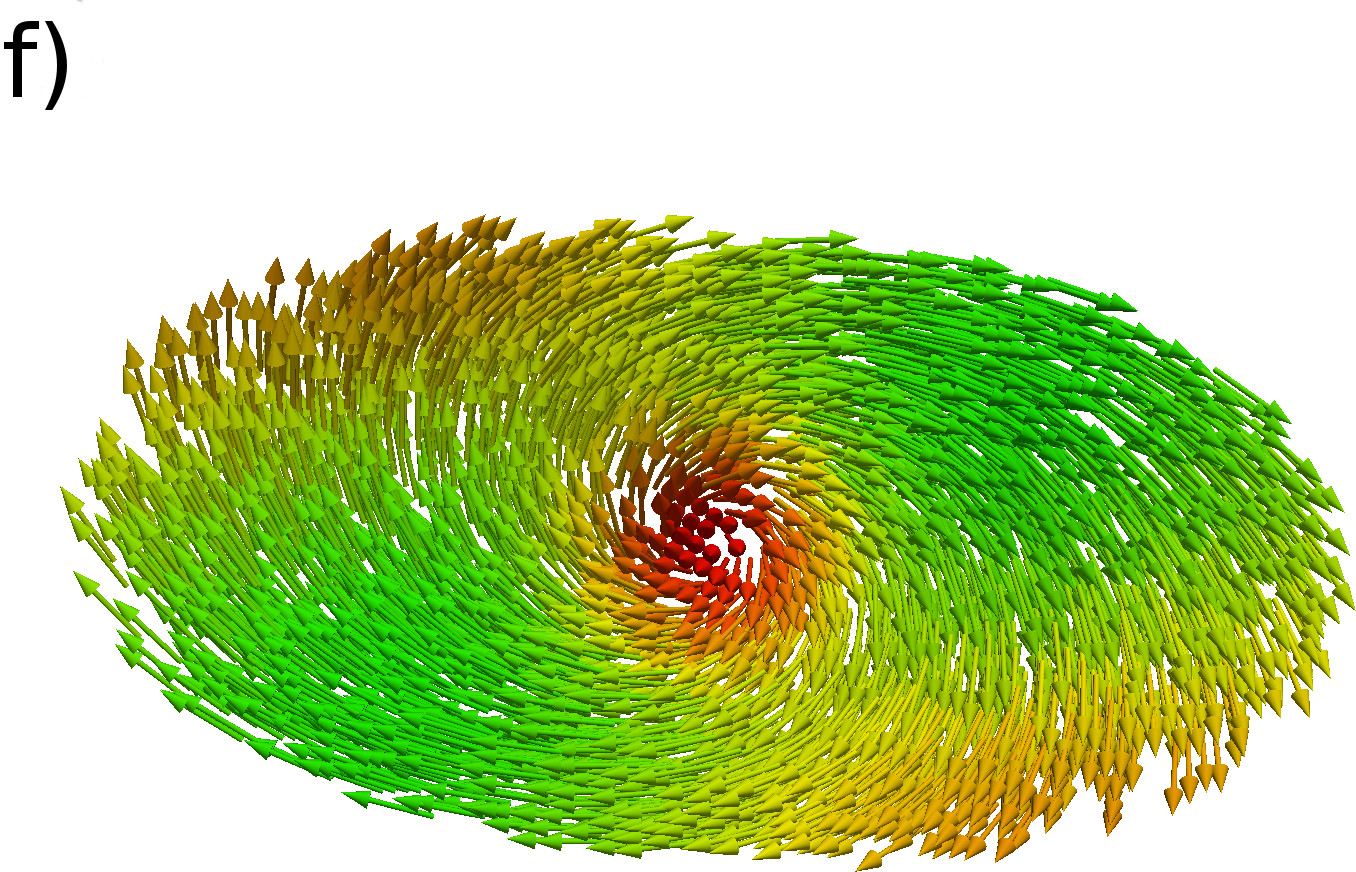}}\caption{Snapshot of the remanent state of the magnetization. Fig.a evidences the presence of a vortex at the external border of the torus. Fig.b consists in a representation of the magnetization in function of $\theta$ and $\varphi$ aiming to show the in-surface component of VA pair. Figs. c, d, e and f show a front view of the planes represented in the Fig.a. These highlighted planes evidence the appearing of a vortex at the external border of the torus (c and f) and the anti-vortex at the internal border (d and e). The horizontal bar shows color scale of the magnetization along $x$-axis.}\label{RemState}
\end{figure}

The LLG equation was solved by using the 3D Object Oriented MicroMagnetic Framework (OOMMF) \cite{oommf-code} package using the computational facilities available in nanohub \cite{oommf-code2}. The simulations were run using an exchange constant $A=1.3\times10^{-11}$ J/m, a saturation magnetization $M_{_S}=860$ A/m, a cubic mesh with size $1$ nm and a damping $\alpha=0.5$. The considered geometric parameters are: toroidal radius $R=52$ nm; internal poloidal radius $r_i=20$ nm; and external poloidal radius $r_e=26$ nm, resulting in a 6 nm thickness torus. In the simulations, we have saturated the magnetization with an external magnetic field in such way that the initial state consists in a single domain along $x$-axis direction. Then, we have diminished the magnetic field until 0 and analyzed the remanent state of the magnetization. 
\begin{figure}
{\includegraphics[scale=0.25]{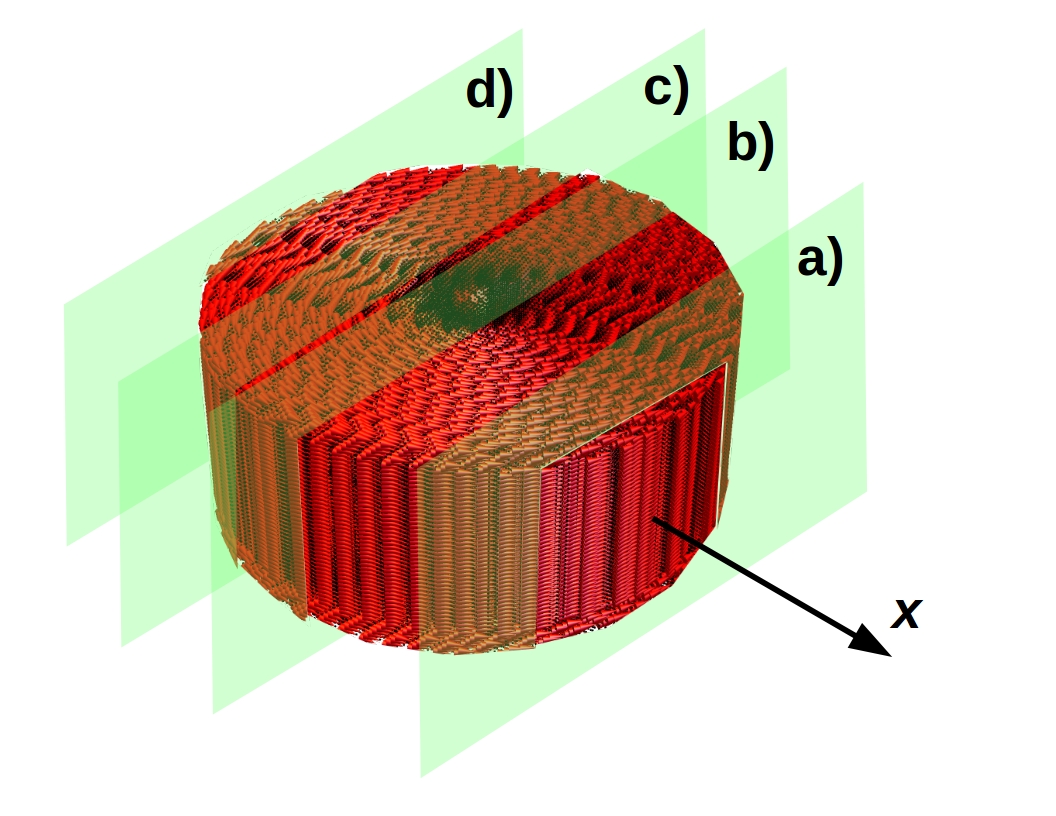}}\\
\includegraphics[scale=0.5]{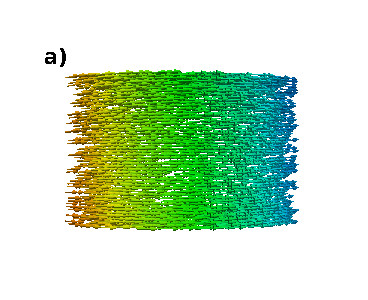}\includegraphics[scale=0.5]{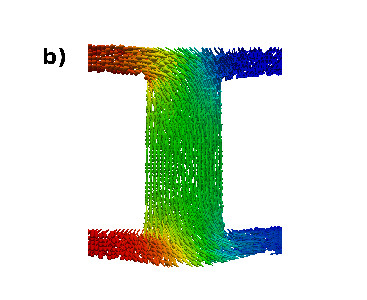}\\\includegraphics[scale=0.075]{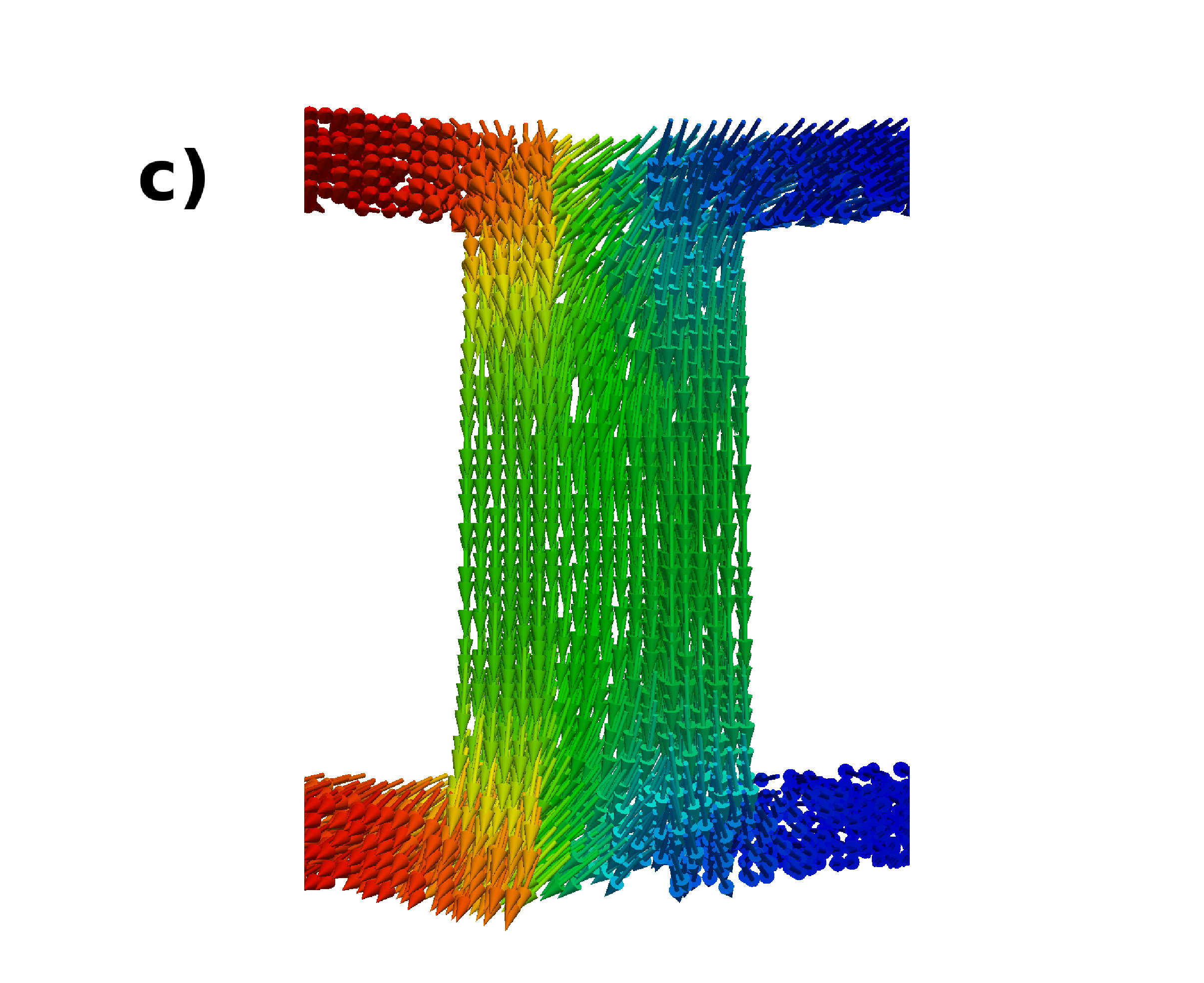}\includegraphics[scale=0.075]{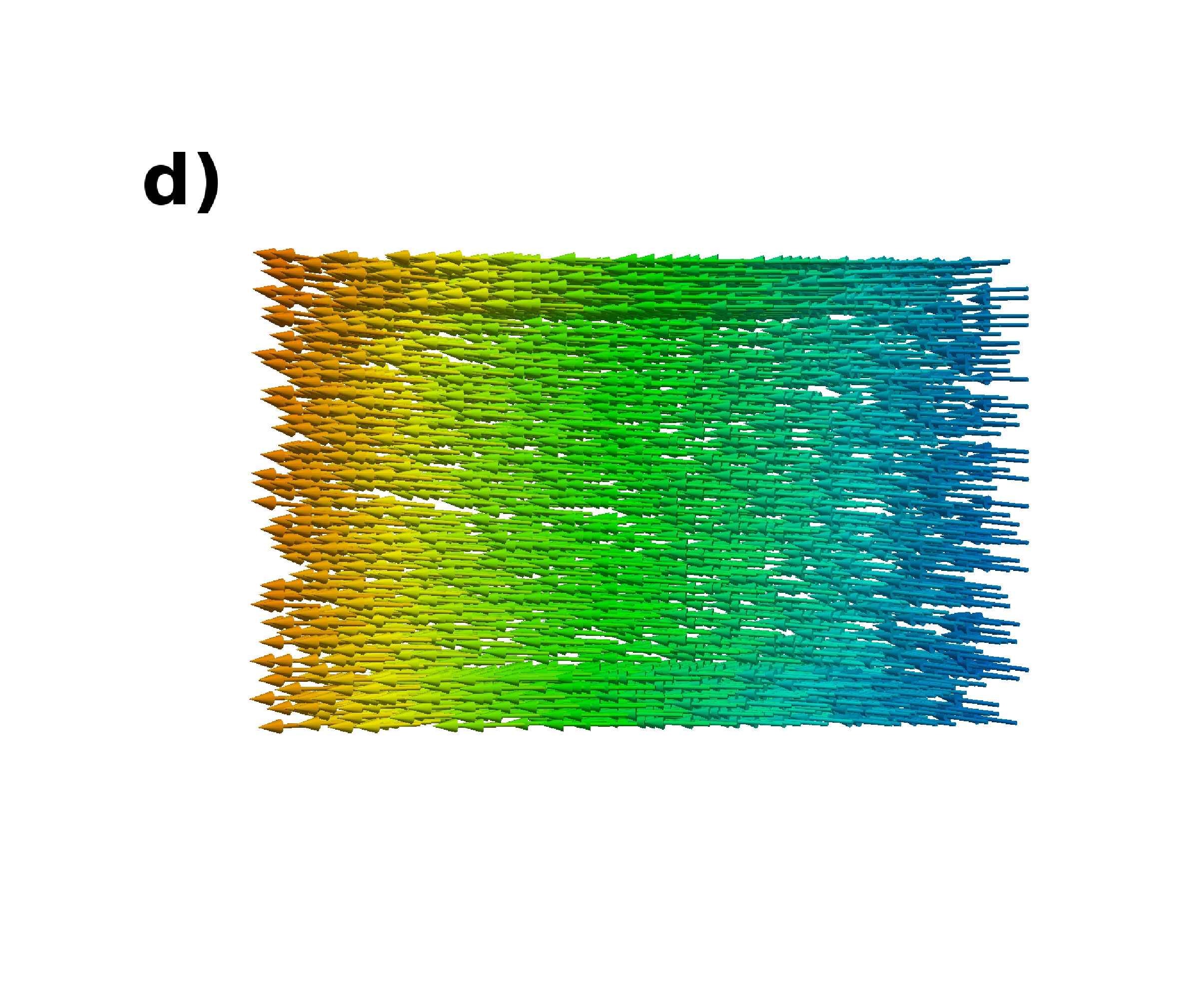}\\
\includegraphics[scale=0.14]{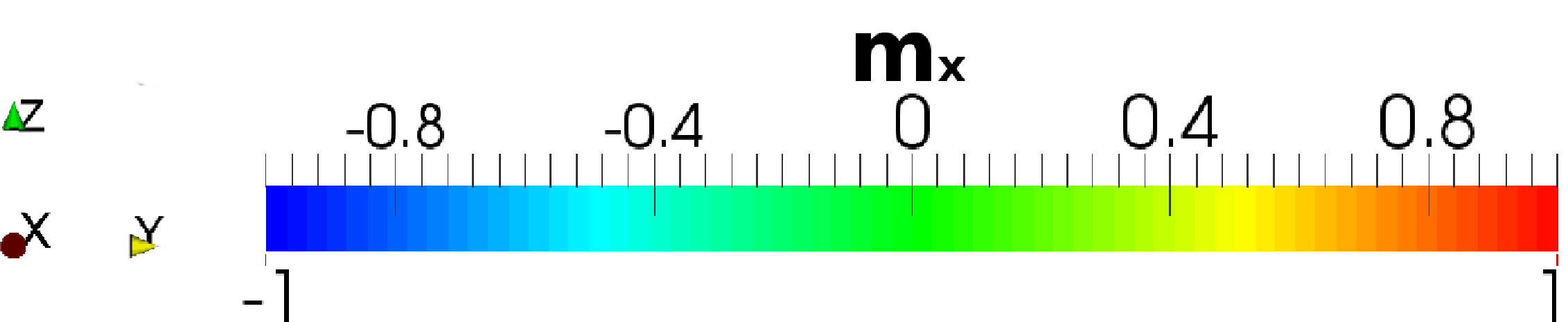}\caption{Snapshot of the remanent state for a holow cylindrical nanoparticle when the magnetic field is pointing along $x$-axis direction. The highlighted planes evidence that no chiral states appear in a hollow cylindrical magnetic particle.}\label{MantoAnular}
\end{figure}

Fig. \ref{RemState}a presents the obtained configuration and the appearance of a vortex at the external torus border is evident. To analyze the magnetization at the internal borders we have done four cutoffs highlighted in the planes showed in Fig. \ref{RemState}a. Fig. \ref{RemState}b shows a representation of the obtained magnetization configuration, given by the in-plane VA ansatz, $\Phi=\text{arg}[r\cos\theta+i(R+r\sin\theta)\cos(\varphi\mp\pi/2)]$. The signals $-(+)$ are related to the interval of the azimuthal angle, that is, $-$ for $\varphi\in[-\pi/4,\pi/4]$ and $+$ for $\varphi\in[3\pi/4,5\pi/4]$. From Figs. \ref{RemState}c-f, it can be observed the formation of two vortices state with opposite chiralities at the external borders (positive curvature) of the torus and antivortices, also with opposite chiralities, at the internal borders (negative curvature) of the torus. Such opposite circulation at the ends of the torus could be associated to the symmetry properties of the LLG equation \cite{Arrot-Book}. That is, if the static magnetic structure is generated by starting from saturation in the in-plane direction and reducing the external field to zero, vortices with opposite chiralities form on opposite ends of a sample. This effect is known, for example, for cylinders or whiskers \cite{Arrot-JMMM} with flat ends that are originally magnetized along their symmetry axis. On the other hand, we have performed micromagnetic simulations for a half torus section with the same previous described geometrical parameters. It has been observed the formation of only one VA pair and there are no two vortices with opposite chiralities in this case. That is, the two vortices appearing at the ends of the integer torus would also appear if the torus would cut on two halves and each half was under the action of an in-plane magnetic field. Then, such pair of vortices at the opposite borders of the torus cannot be associated with the symmetry properties of the LLG equation. Nevertheless, the symmetry of the LLG equation continues to exist because each the VA pair presents opposite chiralities.

These VA pairs can be interpreted as minimal versions of a cross-tie domain wall \cite{Cord-NJP-2009,Middelhoek-JAP-1963}, separating two domains that are formed along the poloidal angle. One domain pointing along $-\hat{\varphi}$ ($-\pi/6\leqslant\theta\leqslant\pi/6$) and another along $+\hat{\varphi}$ ($5\pi/6\leqslant\theta\leqslant7\pi/6$). The formation of one domain in the upper and down regions of the toroidal nanoparticle is a result of the interplay between the shape anisotropy (magnetostatic) and exchange interactions or, in an equivalent way, from the interplay between dipolar, DM-like and curvature-induced effective anisotropy interactions. Indeed, if we consider an in-surface magnetization configuration ($\Theta=\pi/2$) and the previous definitions of $\mathcal{E}_{D}$ and $\mathcal{E}_{A}$ we have that
\ba\label{DMI-torus-Simp}
\mathcal{E}_{D}=\frac{2\,A\,\cos\theta}{(R+r\sin\theta)^2} \partial_\varphi\Phi\,
\ea
and 
\ba\label{Ani-torus-Simp}
\mathcal{E}_A=A\left[\frac{1}{r^2}\cos^2\Phi+\frac{\sin^2\theta}{(R+r\sin\theta)^2}\sin^2\Phi\right]\,.
\ea

From the analysis of Eqs. (\ref{DMI-torus-Simp}) and (\ref{Ani-torus-Simp}), it can be noted that for $\theta=0$ or $\theta=\pi$, $\mathcal{E}_D\sim2/R^2$ and $\mathcal{E}_A\sim1/r^2$. Since Since $R>r$, the solution that minimizes the energy depends on the relation $R/r$. By taking the parameters described in the simulations, we have that $R=2r$ and then $\mathcal{E}_A$ dominates, making the in-surface solution $\Phi=\pi/2$ more favorable. On the other hand, for $\theta\rightarrow\pi/2$, the first term of Eq. (\ref{Ani-torus-Simp}) dominates and it favors the solution $\Phi=0$. Therefore, from the adopted parameters, $\mathcal{E}_D$ even plays the role in such way that spatially inhomogeneous distributions must take place. These spatially inhomogeneous states leads to the formation of observed upper and down domains, evidenced in Fig. \ref{RemState}a. The main consequence of the appearance of these two domains is the formation of the VA pair and in this case, curvature-induced chiral states are observed and DM-like interaction determines the region where the vortex and the antivortex must appear. In fact, by assuming $\rho=\sqrt{(r\cos\theta)^2+[(R+r\sin\theta)\sin\varphi]^2}$, the ansatz
\ba
\Theta=\arccos\left\{\left[1-\left(\frac{\rho}{\rho_c}\right)^2\right]^4\right\}\,,\,\,\,\,\,\,\rho\leq\rho_c
\ea
can be used to describe the vortex (+) and antivortex (-) core with radius $\rho_c$, appearing at $\theta=\pi/2(-\pi/2)$. In this context, the winding number of the metastable state is determined by $\partial_\varphi\Phi$ appearing into the second term of Eq. (\ref{DMI-torus}). Indeed, the signal of second term in Eq. (\ref{DMI-torus}) presents a dependence on $\theta$ in such way that,
\ba
\mathcal{E}_{D^2_{\pm\pi/2}}=\mp\sin\Theta\cos\Theta\cos\Phi\partial_\varphi\Phi\,.
\ea
Therefore, from a direct analogy with the planar case, the ansatz $\Phi=\pm\arctan[r\cos\theta/(R+r\sin\theta)\sin\varphi]$ can be adopted to describe a vortex (+) or antivortex (-) configurations. In this case, above term of the energy is minimized for $\partial_\varphi\Phi<0$ (vortex) around $\theta=\pi/2$ and $\partial_\varphi\Phi>0$ (antivortex) around $\theta=-\pi/2$. Then, the DM-like interaction is responsible for the appearance of a vortex in the external border of the torus and an antivortex in the internal border of the torus. Due to the fact that an antivortex is found at the internal border of the torus, one can conjecture that the antivortex is for a negative curvature as the vortex is for a positive curvature \cite{Spherical-Shell,Vagson-JAP-2015}. Nevertheless, the study on the magnetic groundstate of particles with negative curvature is still lacking.

The antivortex is the topological counterpart of a vortex, having opposite winding number and so, the appearing of two VA pairs preserves the topological winding number of the initial state ($Q=0$ for a single domain). In fact, while a vortex has winding number $Q_{v}=+1$, an antivortex presents $Q_a=-1$. The conservation of the winding number occurs because two configurations belonging to different homotopy classes cannot be continuously deformed one to another \cite{Rajaraman-Book} and thus, isolated vortices (antivortices) or two pairs of vortices (antivortices) are not possible in this case. In addition, the two VA pair state demands lower energy than a two vortices pairs state since if a two pair of vortices would be the remanent state, four antivortices must appear at $\theta=0(\pi)$ $\phi=0(\pi)$ in order to smoothly connect such vortices pairs, increasing the dipolar and exchange energy of the system.  Connecting the VA pairs, it can be noted an in-surface state along the torus that is, the remanent state consists in a quasi-tangential state connecting two VA pairs. From Eq. (\ref{DMI-torus}) and the described geometrical parameters ($R=52$ nm and $r=26$ nm), one can estimate the curvature-induced DM-like interaction effective strength in $\mathcal{D}\sim 0.005 - 0.012$ meV for the external ($\theta=\pi/2$) and internal ($\theta=-\pi/2$) borders of the torus, respectively. 

The formation of VA on a toroidal shell is a very interesting result, both  from the fundamental as well as from the applied point of view. From the fundamental point of view, the appearance of a configuration with positive winding number at the external border and a configuration with negative winding number at the internal border of the torus evidence the intrinsic relationship between curvature and magnetization properties of ferromagnetic nanoparticles. Then, one can conclude that in consequence of a curvature-induced DM-like interaction, there is the possibility of the appearing of a curvature-induced topological metastable state in curved nanomagnets. From the applied point of view, if it is possible to generate and guide the VA state from a curved section of a nanowire to a straight section without annihilate them, this pair could be thought as a candidate to compose devices based on the concept of spin logic operations \cite{Choe-Science-2004} or ``race-track'' memory \cite{Parkin-Science-2008}. Another interesting possibility is to study the interaction of this pair vortex-antivortex with oscillating magnetic fields aiming to use these chiral states as nano-emitter/nano-collector (nanoantenna) devices \cite{Nanoantena1,Nanoantena2}.

To highlight the role of the curvature in the generation of the VA pair, we have performed micromagnetic simulations for a hollow cylinder with same geometrical dimensions of the torus. In fact, differences must be evident when we consider the two geometries, since by parametrizing the cylinder in the natural cylindrical coordinate system ($\hat{n}=\hat{\rho},\hat{q_1}=\hat{\varphi},\hat{q_2}=\hat{z}$), the modified spin connection for the cylindrical geometry is evaluated as $\mathbf{\Omega_c}=0$ and thus, there is not a curvature-induced DM-like interaction term to the magnetic energy in this case (consequently no chiral configuration must be noted). Indeed, by observing the remanent state given in Fig \ref{MantoAnular}, it can be noted that it consists in a single vortex turning around the ring hole and chiral metastable states do not appear in this case. The absence of a VA pair in the sample with exactly the same dimensions but with vanishing curvature is a strong indicator that the origin of the chiral texture lies on the curvature of the hollow torus.

\begin{figure}
\centering
\includegraphics[scale=0.09]{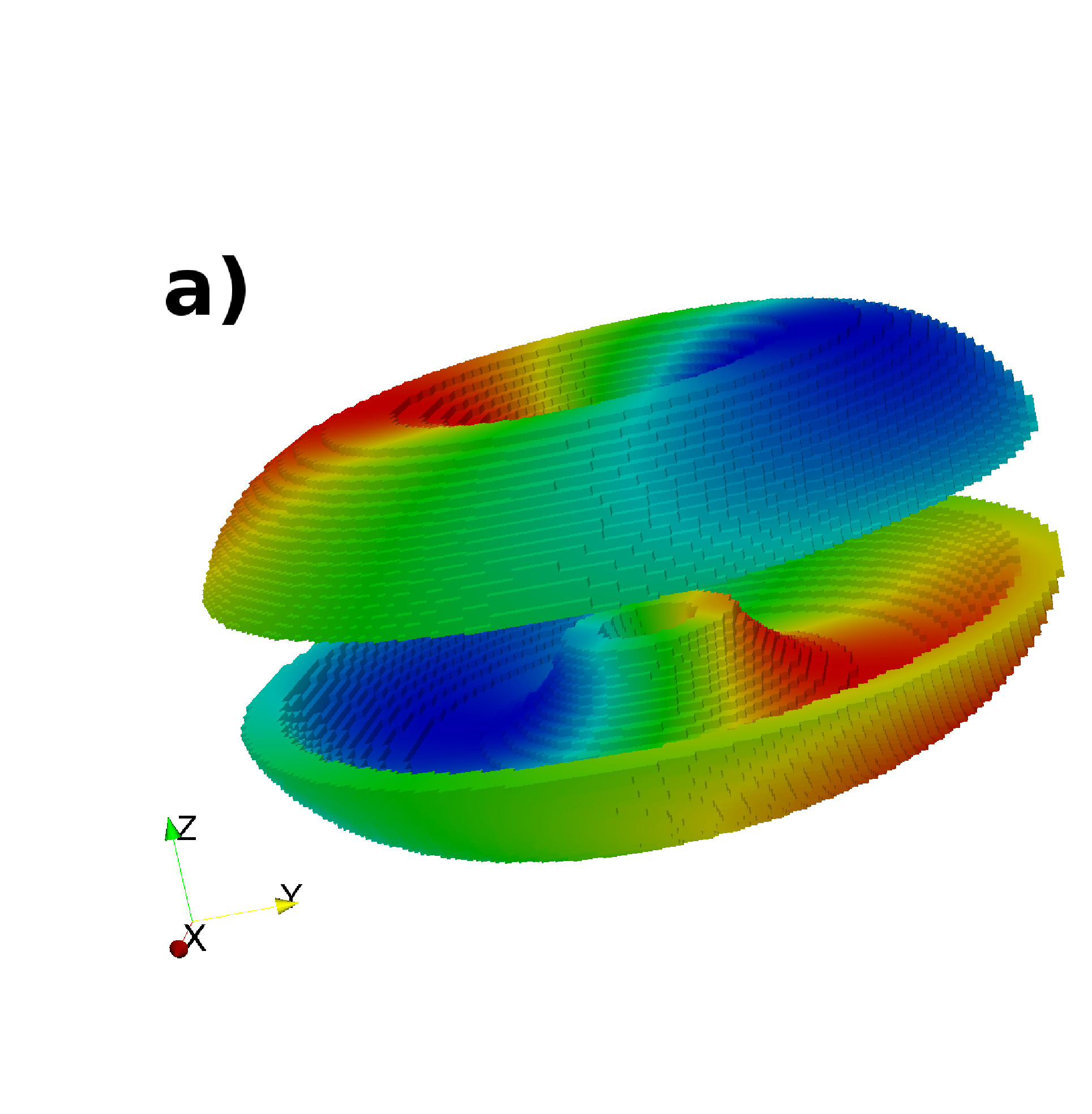}\includegraphics[scale=0.09]{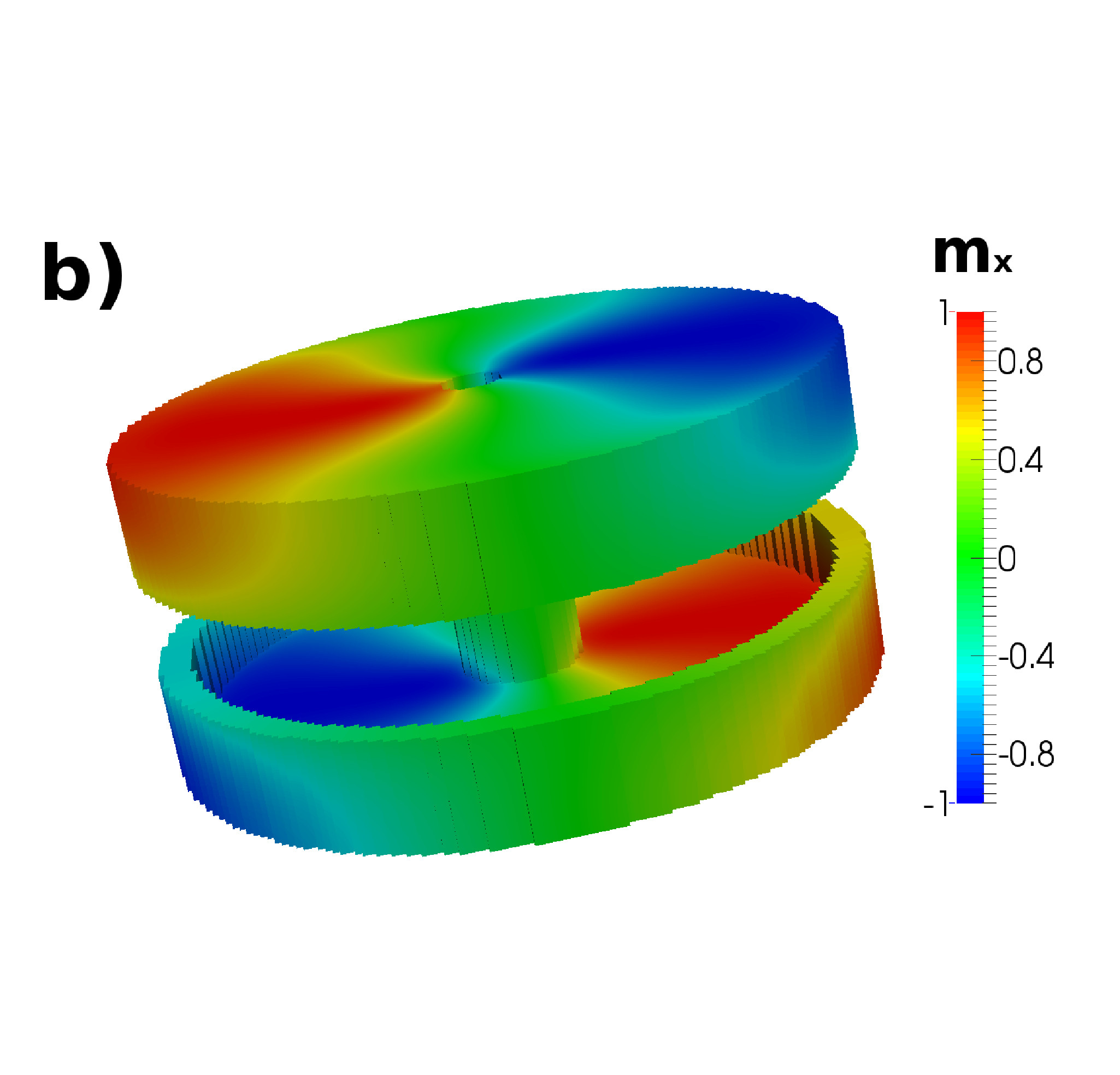}\caption{Remanent states of hollow toroidal (a) and cylindrical (b) nanoparticles when an out-of-plane magnetic field is applied. Vertical bar shows color scale of the magnetization along $x$-axis.}\label{MantoToroidal}
\end{figure}

\begin{figure}
\centering
\includegraphics[scale=0.35]{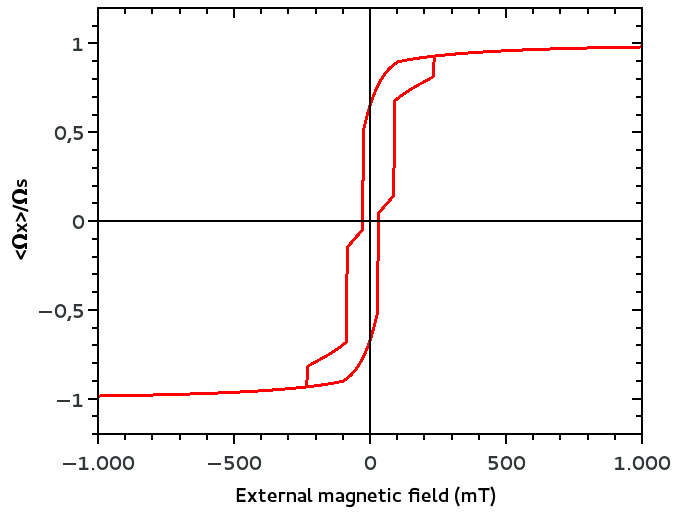}
\includegraphics[scale=0.35]{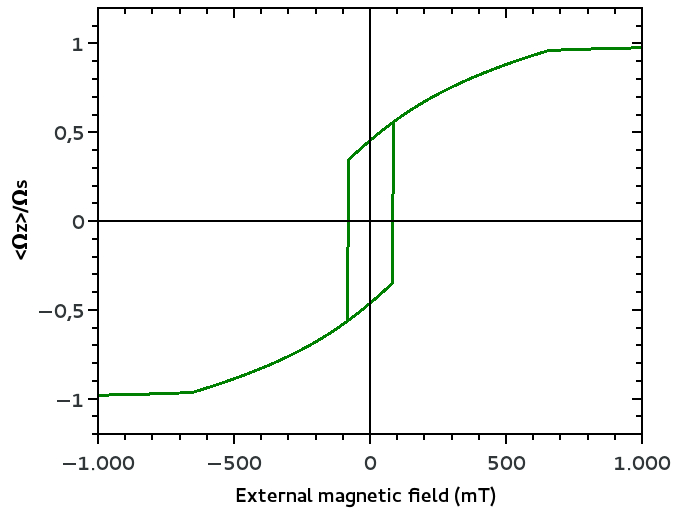}
\caption{Hysteresis curves of reversal processes when an in-plane (a) and an out-of-plane (b) magnetic fields are applied into the system.}\label{RemStateHz}
\end{figure}

On the other hand, if the magnetic field is pointing along the $z$-axis direction, a different behavior must be expected for a hollow cylindrical nanomagnet. In this case the parametrization is given by ($\hat{n}=\hat{z},\hat{q_1}=\hat{\rho},\hat{q_2}=\hat{\varphi}$) and consequently $\mathbf{\Omega_c}=-\hat{\varphi}/\rho$. Therefore, a chiral effect must be noted. In this context, we have also studied the behavior of the magnetization in a hollow toroidal and cylindrical nanomagnets by analyzing the remanent state when an external magnetic field is applied along $z$-axis direction. As expected, due to the symmetry of LLG equation, in  remanence, it can be noted the formation of a double vortex with opposite chiralities for both geometries (See Fig. \ref{MantoToroidal}). However, it can be noted that due to the smooth curvature of the torus, the vortex state is still present at its internal neck while due to the high exchange energy cost to support a vortex, the internal neck of the hollow cylindrical ring presents a single domain state pointing along $z$-axis direction (See Fig. \ref{MantoAnular}b). For both cases, the formation of a two opposite vortex state can be explained by the need to reduce the dipolar energy (vortices with same chirality would lead to larger dipolar energy) and the nanomagnets behave as an array of nanorings separated by a vertical distance $d$ \cite{Escrig-JAP-2006}. This remanent state is also very interesting for applications in spintronic and magnonic devices, due to the possibility to control the chirality of the vortex on one side of the nanomagnet by controlling the other one.

The experimental evidences of the obtained metastable states can be found by analyzing the reversal process of the magnetization during a hysteresis cycle. To understand the mechanism behind the nucleation and annihilation of the observed VA pair ($H_x$) and two opposite vortices ($H_z$) and to show how these states can be experimentally observed, we have studied the hysteresis curves appearing from each reversal processes. The hysteresis curves describing such reversal processes is shown in Fig. \ref{RemStateHz}. It is observed that when the magnetic field is applied along $x$-axis direction, the hysteresis curve presents a reduction in the magnetization at remanence. This reduction evidences the formation of the two VA pairs. From the analysis of a video available in Supplemental Materials \cite{Supp-Mat}, it is observed that as the magnetic field increases in the opposite direction, each VA pair annihilate giving place to a transient state formed by two in-surface domain walls at the opposite sides of the torus. These domain walls joint themselves and the hysteresis curve present a typical neck associated to the nucleation of a vortex state. For $|H_x|\geq100$ mT, the vortex configuration gives place again to a double domain wall, which diminish their lengths when the magnetic field continues to increase, disappearing for $H_x\approx250$ mT. On the other hand, Fig \ref{RemStateHz}b describes the hysteresis curve when the magnetic field is applied along $z$-axis direction. It can be noted a fast decreasing in the magnetization for $H\leq500$ mT, evidencing the formation of the opposite vortices with a small region in which magnetization points along $z$-axis (See supplemental materials \cite{Supp-Mat} for a video). These vortices remain at remanence and are annihilated at $|H_z|\approx100$ mT. It is also noted that in both cases (in-plane and out-of-plane magnetic fields), the magnetic field strength that annihilates the remanent state is small (in the order of 200 mT) and thus, the obtained metastable states are easily nucleated and annihilated. Thus, the creation and annihilation of a VA state in toroidal nanotubes could be used in data storage devices.

\section{Conclusions}
In conclusion, chiral topological interactions induced by curvature can take place in a hollow nanomagnet with variable curvature. Due to its variable curvature, the remanent state of the magnetization in a hollow torus consists in a configuration in which a  vortices appear at the external borders and antivortex are present at the internal border of the torus. Qualitative analysis supports the fact that a VA state has lower energy than a state in which the torus presents two pairs of vortices. From a direct comparison of the results obtained for a geometry with variable curvature with the remanent state of a hollow cylindrical nanomagnet (Gaussian curvature is 0), we showed that the VA state is a result of an effective DMI induced by curvature. In this case, we have shown the new possibility to stabilize vortices and antivortex in magnetic nanoparticle by using curvature. This chiral state could be used in devices working under the concept of spintronic, race-track memory and nanoantennas. Finally, the remanent state when the external magnetic field is pointing along $z$-axis direction consists in a two vortex state with opposite chirality. In this case, main differences between toroidal and cylindrical cases live in the vortex structure in the internal border of both geometries. The reversal processes for both cases ($\mathbf{H}$ pointing along $x$ and $z$ direction) were analyzed. The mechanism behind the appearance and annihilation of the VA pairs have been described by analyzing the hysteresis curves and it was shown that the VA pairs annihilate themselves and the reversal process is followed by the nucleation of a single vortex state.

\textit{Acknowledgements}: We thank the Brazilian agencies CNPq, Fapesb and Fapemig for financial support. ASN would like to thank funding from grant Fondecyt 1150072. ASN also acknowledges support from Financiamiento Basal para Centros Cient\'ificos y Tecnol\'ogicos de Excelencia, under Project No. FB 0807 (Chile). We are gratefull to J. Ot\'alora and A. Bogdanov for their valuable comments on our work.

\end{document}